# Flame Modelling of Premixed Ammonia Combustion in Swirl Burner using LES

Quang Tuyen Le[1], Van Bo Nguyen[1], Shengwei Ma[1], Boon Siong Neo[1], Arthur Lim[1], Chang Wei Kang[1], Huangwei Zhang[2]

[1] Institute of High Performance Computing (IHPC), Agency for Science Technology and Research (A*STAR),
1 Fusionopolis Way, #16-16 Connexis, Singapore 138632, Republic of Singapore
[2] Department of Mechanical Engineering, National University of Singapore, 9 Engineering Drive 1, 117576, Singapore

**ABSTRACT**

The objective of the LES simulation is to investigate how operational parameters affect flame properties in combustion of the premixed $NH_3/H_2$/air inside a swirling burner. Key parameters explored include the differential blend ratio of $NH_3/H_2$ fuels, ignition conditions, flow dynamics, and the balance between accuracy and computational cost using reduced chemical kinetic mechanisms. The numerical simulation findings suggest optimal operational conditions for enhancing thermal efficiency and controlling NOx emissions for experimental setups, scaling up combustion platforms.

## 1. Introduction

Ammonia has recently received great attention from researchers and engineers as it presents various advantages over other fuels as a hydrogen carrier as well as ammonia/hydrogen blends [1,2]. However, it requires numerous advanced research efforts to improve the combustion efficiency before utilizing ammonia in gas turbines applications [2,3] because of low heat combustion, laminar burning speed, and flammability in ammonia/air flames. In addition, burning $NH_3$ emits undesired $NO_x$ and $NH_3$ slip which is dependent on combustion regime. Both experiment and numerical studies show that both efficiency and species emissions are sensitive to combustion operating conditions and there are rooms to improve combustion properties further to achieve the stage of utilizing ammonia in turbine application. For example, experimental study showed that flames burning $NH_3/H_2/N_2$/air mixture are more durable to blow off than $CH_4$/air composition in an axisymmetric unconfined bluff-body stabilized burner [4]. The measurement of NO species suggests that introducing hydrogen into the mixture can cause a shift of $NO_x/NH_3$ emissions trade-off toward the rich side [5].

Accompanying to these expensive experimental investigations, the advanced numerical computational methods and computer recourse help to explore insights into the chemical and physical phenomena within the combustion chamber as well as test new proposals or concepts for gas turbine. A Large Eddy Simulation (LES) simulation can predict quite accurately NOx emissions in swirler burner using a ratio 1:1 of $NH_3/H_2$ fully re-mixed fuel with a lean equivalence ratio range from 0.46 to 0.56 [6]. The premixed fuel in swirler burner has been focused on much research as it shows some optimal points in burning process. The minimum of NOx and $NH_3$ emission can be achieved at condition of equivalence ratio of 1.225 and initial mixture temperature of 500K in LES simulation of premixed $NH_3$/air flame in swirler burner [7]. Still, numerical simulations of reacting flow largely rely on kinetics modelling which describes chemical reaction inside combustion chamber and numerical method accuracies.

In this work, utilizing the OpenFOAM platform, we conduct high-fidelity Large Eddy Simulation (LES) of premixed ammonia combustion in a swirl burner. Our aim is to investigate critical operational parameters, such as the differential blend ratio of $NH_3/H_2$ fuels, ignition conditions, and flow dynamics. The numerical results will provide insights into optimal operational conditions and chamber configurations for subsequent experimental studies.

## 2. Methodology
### 2.1 Configuration of Combustor and Operating Conditions.

The laboratory-scale model with the chamber size of $1250 cm^3$ is used in this study. The premixed fuel is swirler by 8 radical canals inside the cylinder tube before entering the chamber zone. The cubic chamber is connected by converging cone and exit pile, shown in Fig.1.

The fully premixed fuel as component of 99.99% purity ($NH_3$, $H_2$ and synthetic air) is considered in this study. The room condition (T= 300K and P = 1atm) is applied for inlet condition. The inlet mass flowrate is fixed at 0.0205kg/s meanwhile the ratio of fuel/air in

Table 1 Combustion chamber operating conditions.

| Parameter | Value |
| --- | --- |
| P [atm] | 1 |
| $T_{inlet}$ [K] | 300 |
| Flowrate [kg.s$^{-1}$] | 0.0205 |
| Equivalent ratio | 0.8-1.3 |

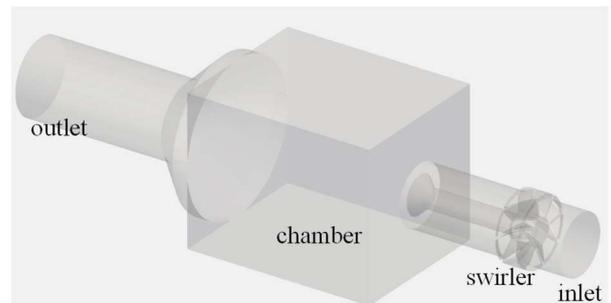

Fig. 1 Simulated geometry of swirler burner.

---

Corresponding author: Quang Tuyen Le
*E-mail address*: leqt@ihpc.a-star.edu.sg

mixture can be varied in range 0.8 to 1.3 as summarized in Table 1.

**2.2 Numerical Set Up.**

The flow inside combustor is solved by LES model which directly calculates large-scale turbulent structure and small structure are modeled by filtering. The governing equation are equations are treated with Favre-filtered as:

$$\frac{\partial \bar{\rho}}{\partial t} + \frac{\partial \bar{\rho} \tilde{u}_j}{\partial x_j} = 0 \qquad (1)$$

$$\frac{\partial \bar{\rho} \tilde{u}_i}{\partial t} + \frac{\partial (\bar{\rho} \tilde{u}_i \tilde{u}_j + \bar{p} \delta_{ij})}{\partial x_j} = \frac{\partial (\tilde{\tau}_{ij} - \tau_{ij}^{sgs})}{\partial x_i} \qquad (2)$$

$$\frac{\partial \bar{\rho} \tilde{e}_t}{\partial t} + \frac{\partial (\bar{\rho} \tilde{e}_t \tilde{u}_j + \bar{p} \tilde{u}_j)}{\partial x_j} = \frac{\partial (-\bar{q}_i + \tilde{u}_j \tilde{\tau}_{ij} - Q_i^{sgs} + \sigma_i^{sgs} - H_i^{sgs})}{\partial x_i} \qquad (3)$$

$$\frac{\partial \bar{\rho} \tilde{Y}_i}{\partial t} + \frac{\partial (\bar{\rho} \tilde{u}_i \tilde{Y}_j)}{\partial x_j} = \frac{\partial (Y_j^{sgs} + \bar{\rho} \tilde{D}_i \tilde{Y}_j)}{\partial x_i} + \bar{\omega}_i \quad i=1..N_s \qquad (4)$$

The equations 1 to 3 describe mass, momentum and energy while equation 4 describes a species transportation. In those equations, the overbar denotes the spatial-filtering operation, while a tilde is the Favre-filtering operation. The variables u, t, ρ, p, e, τ, q and $\tilde{Y}_j$ represent the velocity, time, density, pressure, total energy, viscous stress, heat flux and mass fraction of species, respectively. The terms $\sigma_i^{sgs}$, $H_i^{sgs}$, $\tau_{ij}^{sgs}$, $Q_i^{sgs}$ and $Y_j^{sgs}$ are the subgrid-scale terms, which are treated using the Smagorinsky model for compressible flow. $\bar{\omega}_i$ is production rate due to reactions, which is evaluated by solving the system of stiff ODE equations for Ns species. In addition, the detailed chemical kinetic model which including the species and involving equation, is employed to describe the combustion reaction progress. In this study, the trade-off of accuracy and cost of differential reduced chemical kinetic models is explored, starting with the first model includes 21 species and 122 reactions.

The compressible PIMPLE solver in OPENFOAM 2112 is employed to deal with the strong coupling between velocity and pressure, scalar equations. The maximum Courant–Friedrichs–Lewy (CFL) number is fixed at 0.2 (maximum time step ~ 1 μs) to ensure the stability and accuracy of simulation. The mesh sensitivity is conducted through 4 meshes from 1.5 to 12 million cells, corresponding to the mesh resolution decreases from 1mm to 0.5 mm.

The simulation includes two steps: At first, the flow passes throughout the entire domain without chemical reaction, resulting in steady state for fluid and distribution of inlet species. Then, the combustion process with kinetics reaction is modeled by activating chemical reaction with ignition temperature (T ~ 1500K) near the entrance of chamber.

**3. Preliminary Results and Discussion**

The iso-surface contours of OH species mass fraction (magnitude from 0.0035 to 0.006) colored by temperature is shown in Fig.2. As involving in NO forming, autoignition dynamic and de-NOx process of combustion, OH radical plays rule role to help us understanding combustion kinetics, pollutant formation, and flame stability. The OH is formed at diffuser nozzle of inlet pile and concentrates near the exit of chamber, while is almost empty at corners of chamber.

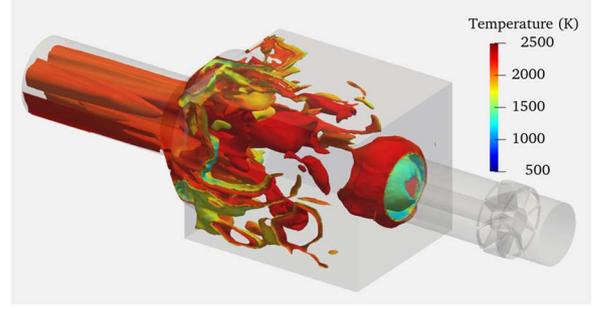

Fig. 2 Iso-surface of OH species, colored by temperature.

The contour of temperature shows that the $NH_3$ combustion inside the chamber can release heat above 2000K, shown in Fig.3. It also shows that there are some locations where the burning $NH_3$ is not completed yet. It is visually presented by contours of $NH_3$ in the next column: there is a correlation position where low temperature and high concentration of $NH_3$. The NO species represents almost inside the chamber, specifically mass fraction of NO is up to 0.02 near the entrance of chamber. This suggests that this operation condition (equivalence ratio of 1 and flowrate of 0.0205kg/s) is not optimal for mitigating the pollutant emission.

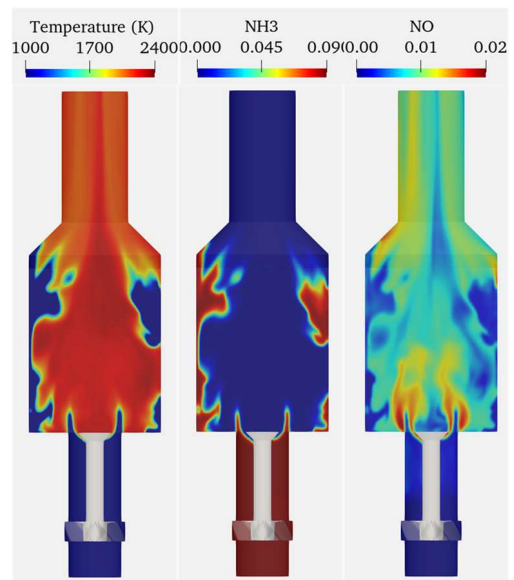

Fig. 3 Contours of temperature, $NH_3$ and NO at the center plane of chamber.

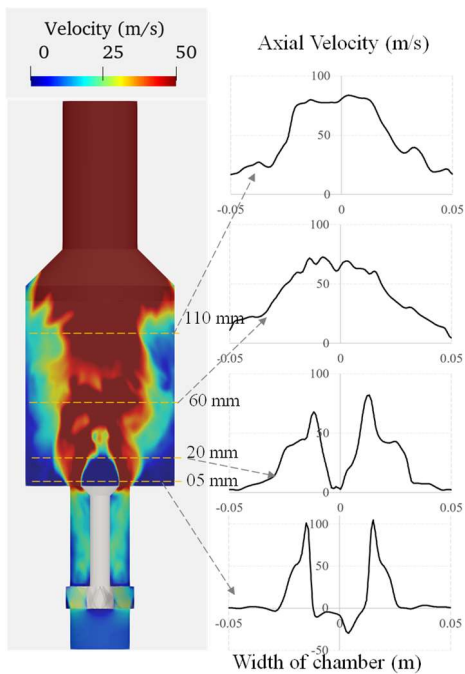

Fig. 4 Contours of axial velocity at central plane and its' profile at differential axial locations.

The Fig. 4 shows the axial velocity contours and its profile at differential axial locations starting from the entrance of chamber. Axial velocity shows the stability of flame and percentage of fuel oxidation. Near the entrance of the chamber (h = 5mm), the inlet velocity is accelerated up to 3 times compared to that of the case without reaction. The stagnation zone at the center behind the blunt tube extends up to 20 mm along the axial direction. The flame structure seems to stabilize from 60 mm to downstream.

## 4. Concluding Remarks

This study presents a Large Eddy Simulation (LES) of premixed ammonia combustion in a laboratory-scale swirler burner under varying operating conditions. The results indicate that both the $NH_3/H_2$ ratio and the fuel/air ratio significantly influence flame properties, thermal efficiency, flame stability, and the combustion process's flammability.

Simulated data on species concentrations, temperature profiles, and flame velocities are analyzed to provide a deeper understanding of the combustion dynamics. The findings reveal that, for a fixed NH3/H2 ratio, an optimal fuel/air ratio (approximately 1.1) maximizes thermal efficiency compared to leaner or richer fuel conditions. However, further investigation is necessary to determine the optimal operational conditions for enhancing combustion efficiency and controlling NOx emissions.

**Acknowledgment**

This work is supported by the RIE2025 USS LCER PHASE 2 PROGRAMME HETFI DIRECTED HYDROGEN PROGRAMME under Grant ID U2305D4002.